\documentclass[%
 aip,
 amsmath,amssymb,
 reprint,%
]{revtex4-1}

\usepackage{graphicx}
\usepackage{dcolumn}
\usepackage{bm}

\usepackage[utf8]{inputenc}
\usepackage[T1]{fontenc}
\usepackage{mathptmx}
\usepackage{etoolbox}
\usepackage{graphicx}
\usepackage{cleveref}
\crefname{figure}{Fig.}{Figs.}
\crefname{equation}{Eq.}{Eqs.}
\makeatletter
\def\@email#1#2{%
 \endgroup
 \patchcmd{\titleblock@produce}
  {\frontmatter@RRAPformat}
  {\frontmatter@RRAPformat{\produce@RRAP{*#1\href{mailto:#2}{#2}}}\frontmatter@RRAPformat}
  {}{}
}%
\makeatother
\begin{document}

\preprint{AIP/123-QED}

\title{The evolution of cooperation in spatial public
goods game with tolerant punishment based on
reputation threshold}
\author{Gui Zhang}
\affiliation{The College of Artificial Intelligence, Southwest University, No.2 Tiansheng Road, Beibei, Chongqing, 400715, China.}
\author{Yichao Yao}
\affiliation{The College of Artificial Intelligence, Southwest University, No.2 Tiansheng Road, Beibei, Chongqing, 400715, China.}
\author{Ziyan Zeng}
\affiliation{The College of Artificial Intelligence, Southwest University, No.2 Tiansheng Road, Beibei, Chongqing, 400715, China.}
\author{Minyu Feng\textsuperscript{*}}
\affiliation{The College of Artificial Intelligence, Southwest University, No.2 Tiansheng Road, Beibei, Chongqing, 400715, China.}

\email{myfeng@swu.edu.cn}
\author{Manuel Chica}
\affiliation{Department of Computer Science and A.I. Andalusian Research Institute DaSCI “Data Science and Computational Intelligence", University of Granada, 18071, Granada, Spain.}
\affiliation{School of Information and Physical Sciences, The University of Newcastle, NSW 2308, Callaghan, Australia.}
\date{\today}

\begin{abstract}
Reputation and punishment are significant guidelines for regulating individual behavior in human society, and those with a good reputation are more likely to be imitated by others. In addition, society imposes varying degrees of punishment for behaviors that harm the interests of groups with different reputations. However, conventional pairwise interaction rules and the punishment mechanism overlook this aspect. Building on this observation, this paper enhances a spatial public goods game in two key ways: 1) We set a reputation threshold and use punishment to regulate the defection behavior of players in low-reputation groups while allowing defection behavior in high-reputation game groups. 2) Differently from pairwise interaction rules, we combine reputation and payoff as the fitness of individuals to ensure that players with both high payoff and reputation have a higher chance of being imitated. Through simulations, we find that a higher reputation threshold, combined with a stringent punishment environment, can substantially enhance the level of cooperation within the population. This mechanism provides deeper insight into the widespread phenomenon of cooperation that emerges among individuals.
\end{abstract}

\maketitle

\begin{quotation}
In the real world, individual behavioral choices have attracted a large amount of attention. In complex networks, cooperation and defection of nodes reflect real-life behaviors. Edges generate interactions, and spatial public goods games (SPGG) help to understand social choices. Realistic social individuals are more likely to choose bad behaviors due to the high rewards of free-riding behaviors, and reputation and punishment mechanisms are crucial to behavioral regulation. A good reputation leads to trust and cooperation, while a bad reputation leads to isolation. In addition, real societies have an assessment of individual reputations, resulting in different communities and corresponding management systems. This paper explores the evolutionary punishment mechanism of reputation and tolerance, combining reputation and punishment to construct a community mechanism that promotes cooperation, aiming to reveal behavioral patterns and provide guidance for social cooperation.

\end{quotation}

\section{Introduction}\label{sec1}

The study of cooperation in biology, sociology, and economics is extensive \cite{perc2016phase,sachs2004evolution,archetti2011economic,axelrod1981emergence,feng2024information,li2024asymmetrical}. When confronted with conflicts between individual interests and general interests, the selfish choices of individuals often clash with the public interests of the group, thereby hindering the emergence of cooperation \cite{ludsin2001biological,hauser2014cooperating,rand2013human,clutton2009cooperation}. Consequently, evolutionary game theory has attracted significant attention and extensive research. Prisoner dilemma game (PDG) \cite{zeng2022spatial,yang2024interaction,zhang2024cooperative,jia2024freedom}, snow drift game (SDG) \cite{zhang2020oscillatory,hauert2004spatial,souza2009evolution,zhu2017evolutionary}, and public goods game (PGG) 
\cite{li2024conditional,santos2008social,hauert2002volunteering} are used to address dilemmas observed in evolutionary games \cite{gintis2000game,perc2010coevolutionary}. The resolution of the social dilemma is noteworthy, with numerous studies proposed to elucidate the emergence of cooperation. For example, Nowak and May \cite{nowak1992evolutionary} are the first to show that spatial structure can promote the development of cooperation in the PDG in 1992. Since then, substantial research on cooperative behavior has been carried out in various evolutionary game models, including those based on square grids \cite{li2021evolution}, small world networks \cite{deng2010memory,fu2007evolutionary,masuda2003spatial,szabo2007evolutionary}, scale-free networks \cite{kleineberg2017metric,wu2007evolutionary,nowak2004evolutionary,dui2020analysis}, and hypergraphs \cite{civilini2021evolutionary,wang2024evolutionary, xu2024reinforcement}. Considering that the previously mentioned network models neglect the typical social phenomenon of temporarily offline vertices and the need to account for hidden vertices, as well as the switching of vertex phases between online and hidden states, Zeng's work on temporal network modeling with online and hidden vertices, based on the birth-and-death process, provides a framework for evolutionary games \cite{zeng2023temporal}.

As researchers strive to elucidate the dilemma, various mechanisms that facilitate cooperation are investigated. Nowak \cite{nowak2006five} put forward five cooperative evolutionary rules, namely kin selection, direct reciprocity, indirect selection, group selection, and network reciprocity. Subsequently, with the rapid advancement of social network theory, numerous mechanisms, including rewards \cite{motepalli2021reward,szolnoki2010reward}, punishment \cite{podder2021reputation}, and social exclusion \cite{szolnoki2017alliance,liu2019evolutionary} are thoroughly studied in recent decades. For example, Liu incorporates both prosocial and antisocial exclusion strategies related to public pools into the public goods game, examining their effects on cooperation. The findings indicate that social exclusion proves to be more effective in reducing defection and fostering cooperation compared to costly punishment \cite{liu2020evolutionary}. In addition, Liu demonstrates that unequal status significantly influences cooperative behavior, with smaller nodes more likely to engage in positive cooperation, thus providing a supplementary perspective on the aforementioned mechanisms \cite{liu2021understanding}.

In contrast to biological systems, human societies possess a credit system in which reputation plays a crucial role, particularly in business collaborations. Since Nowak and Sigmund \cite{nowak1998evolution} propose the image-scoring reputation model, reputation-based mechanisms are shown to effectively enhance cooperation in various social dilemmas \cite{podder2021reputation,milinski2002reputation}. In recent years, Xia \textit{et al.} review the research related to reputation, with a focus on image scores as the bearers of reputation and various forms of reciprocity including direct, indirect, and network reciprocity, demonstrating how these definitions influence the evolution of cooperation in social dilemmas \cite{xia2023reputation}. In the context of cooperation, reputation can influence an individual's behavior and decisions. For example, individuals with a good reputation are more likely to be trusted and cooperate with others \cite{quan2020reputation, feng2024evolutionary}. Punishment is also one of the important mechanisms for regulating individual behavior. It is another important aspect of the promotion of cooperation. When individuals face punishment for non-cooperative behavior, they are more likely to choose cooperation \cite{yamagishi1986provision,ostrom1992covenants,fehr2000cooperation}. However, there is no agreed-upon norm for the mechanism of individual punishment. Some studies prove that expensive punishment is not applicable, as well as its role in rewards \cite{rand2009positive} and the imposition of antisocial punishment \cite{herrmann2008antisocial,rand2010anti,gachter2011limits,rand2011evolution}. Furthermore, reputation and punishment mechanisms are confirmed to play a particularly important role in influencing the ability of groups to participate in effective collective actions. For example, Brandt integrates reputation and punishment mechanisms into the PGG. By studying the complex situation of interaction among three individuals, he demonstrates the promoting effects of territoriality, punishment opportunities, and reputation on cooperative behavior. He reveals the interesting phenomenon that less cooperative individuals can create a more cooperative society \cite{brandt2003punishment}.
Furthermore, Podder finds that choosing social norms that impose more moderate reputation-based punishment will increase cooperation \cite{podder2021reputation}. In general, effectively combining these two mechanisms to promote cooperation in groups remains one of the difficulties in current research.

In community development projects, public facility construction projects are similar to PGG. Individuals can choose to cooperate or defect. Due to the high returns brought about by defecting, individuals often tend to choose this option. Several studies show that the addition of punishment or reputation mechanisms can significantly promote the emergence of cooperation. Inspired by the tolerance mechanism for defectors in a harsh environment \cite{quan2020reputation} and considering community development, if a resident group actively participates in community volunteer services for a long time and has a good reputation, then even if the contribution of individual members is slightly lower than expected, they will not be severely punished. This reflects the concept of tolerance for high-reputation groups.

Therefore, this paper focuses on the combination of reputation and punishment mechanisms. We set a reputation threshold for the incorporation of punishment in the SPGG, with a third-party punishment institution showing tolerance to defectors when the average reputation of a game group is above this threshold. Our model has two major improvements: one is a reputation evaluation mechanism that allows individuals in high-reputation groups to defect without corresponding punishment. The second is the use of fitness functions based on reputation. 

Simulations in square lattices show that the tolerance and punishment mechanism based on average reputation promotes cooperation. In addition, a larger reputation threshold and punishment can create a better cooperation environment.

The remainder of this paper is organized as follows. The reputation-based tolerant punishment PGG model is elaborated in Section \ref{model}, including the reputation evolution, punishment, and strategy update rules. Subsequently, the specific process and details of the simulation, along with the results and discussions of the experiment, are presented in Section \ref{simulation}. Finally, we summarize the entire context in Section \ref{outlook}.

\section{Model}\label{model}

\subsection{Spatial public goods game}

SPGGs are vital for analyzing the cooperative dilemma present in spatially structured populations. This model involves multiple participants and operates as a two-strategy game. In an SPGG, each player occupies a node in the network and engages with their neighboring participants. Let $N(i)$ denote the set of direct neighbors of the player $i$. At each stage, the player $i$ participates in the $N(i) + 1$ groups of the SPGG simultaneously. One group revolves around the player $i$ itself, while the other $N(i)$ groups involve each of its neighboring participants.

In these multi-group games, the individual $i$ must decide whether to contribute one unit to the public pool. In each group game, all contributions to the public pool are multiplied by an enhancement factor \(r\) (\(r > 1\)). The resulting total is then equally distributed among all participants in that group. Let \(s_i\) denote the strategy chosen by the player \(i\). In this framework, the player \(i\) has two choices: cooperation (\(s_i = 1\)) or defection (\(s_i = 0\)). In the \(g\)-th group game (\(1 \leq g \leq N(i) + 1\), with the set of vertices of this group denoted as \(G_g\)), the payoff for player \(i\) as cooperator, denoted as \(\Pi_g^C\) and as defector, denoted as \(\Pi_g^D\), is defined as \cref{1}.

\begin{equation}
\begin{cases}
\Pi_g^C = r\frac{(N_g^C + 1)}{|G_g|}-1\\

\\\Pi_g^D = r\frac{N_g^C}{|G_g|}
\end{cases}. \label{1}
\end{equation}

Here, \(N_g^C\) indicates the count of cooperators, in addition to player \(i\), within their \(g\)-th group game. Consequently, the total payoff \(\Pi_i\) for individual \(i\), who participates in \(N(i) + 1\) group games, can be calculated as \(\Pi_i = \sum_{g = 1}^{N(i) + 1} \Pi_g^{s_i}\).

\subsection{Reputation assessment of the players}

We incorporate reputation assessment into the SPGG. In this subsection, all players receive reputation scores over time, $R_i(t)$, based on their actions within the evolutionary game. Additionally, when a player contributes to the public pool, their reputation increases by one unit during that time step. In contrast, if a player does not contribute, their reputation will decrease by one unit, as defined in \cref{2}. For simplicity and to avoid an unbounded increase in individual reputations, we restrict the reputation values to a range of [0, 20]. If a player's reputation surpasses 20, it will be capped at this maximum value. Conversely, if a player's reputation falls below 0, it will be adjusted to remain at 0.

\begin{equation}
    R_{i}(t)=
\begin{cases}
R_{i}(t - 1)+1, &\text{if } s_{i}(t)=1\\
R_{i}(t - 1)-1, &\text{if } s_{i}(t)=0
\end{cases}. \label{2}
\end{equation}

\subsection{Punishment based on low reputation individuals of the group}

Furthermore, we incorporate a punishment mechanism into the SPGG to encourage cooperation rather than defection. Based on dynamic reputation, each group has a distinct average reputation. We set a reputation threshold $R_{0}$. In the $k$-th group game, if the average reputation of the players is less than $R_{0}$, a third-party institution will punish $b$, being $0 \le b \le 1$, on defectors. Therefore, the benefit function in a low-reputation group changes from \cref{1} to the following \cref{3} .

\begin{equation}
\begin{cases}
\Pi_g^C = r\frac{(N_g^C + 1)}{|G_g|}-1\\

\\\Pi_g^D = r\frac{N_g^C}{|G_g|}-b
\end{cases}. \label{3}
\end{equation}

\subsection{Strategy evolution}

In this way, we take the reputation into account in the update of the strategy. In order to be more in line with real society, we determine that individuals with higher reputation and higher payoff have higher fitness. In the model, individuals with a reputation higher than the reputation threshold will have a higher fitness. In contrast, these individuals with a reputation lower than the reputation threshold will have a lower fitness. We adopt a simple linear weighted method and use a damping factor $\delta$ to adjust the relative importance of these two factors. Therefore, we evaluate the fitness of the agent $i$ according to \cref{Eq:(3)}.

\begin{equation}
f_{i}(\Pi_{i},R_{i})=\delta\Pi_{i}+(1-\delta)\frac{R_{i}-R_{0}}{\lambda}.\label{Eq:(3)}
\end{equation}

In this paper, we set $\lambda$ as the maximum reputation of 20 to ensure that the difference in fitness brought by reputation is fixed between 0 and 1. Next, the probability that agent $i$ adopts the strategy of agent $j$ in the upcoming round of the game is determined by
\begin{equation}
    \Gamma\left(S_i\to S_j\right)=\frac{1}{1+e^{\frac{f_j\left(\Pi_jR_j\right)-f_i\left(\Pi_iR_i\right)}{\kappa }}}.\label{eq:4}
\end{equation}

Here, $f_j\left(\Pi_jR_j\right)$ and $f_i\left(\Pi_iR_i\right)$ denote the fitness of player $j$ and player $i$, respectively. $\kappa$ ($\kappa > 0$) is a noise parameter that represents the uncertainty in the updating of the strategy. When $\kappa\rightarrow0$, if the player $j$ has a higher fitness, the player $i$ will definitely imitate the player $j$. When $\kappa\rightarrow\infty$, whether the player $i$ imitates the player $j$ is completely random. Based on other scholarly works, we set $\kappa = 0.5$ in this study.

\begin{figure}[h!]
	\centering
		\includegraphics[scale=.45]{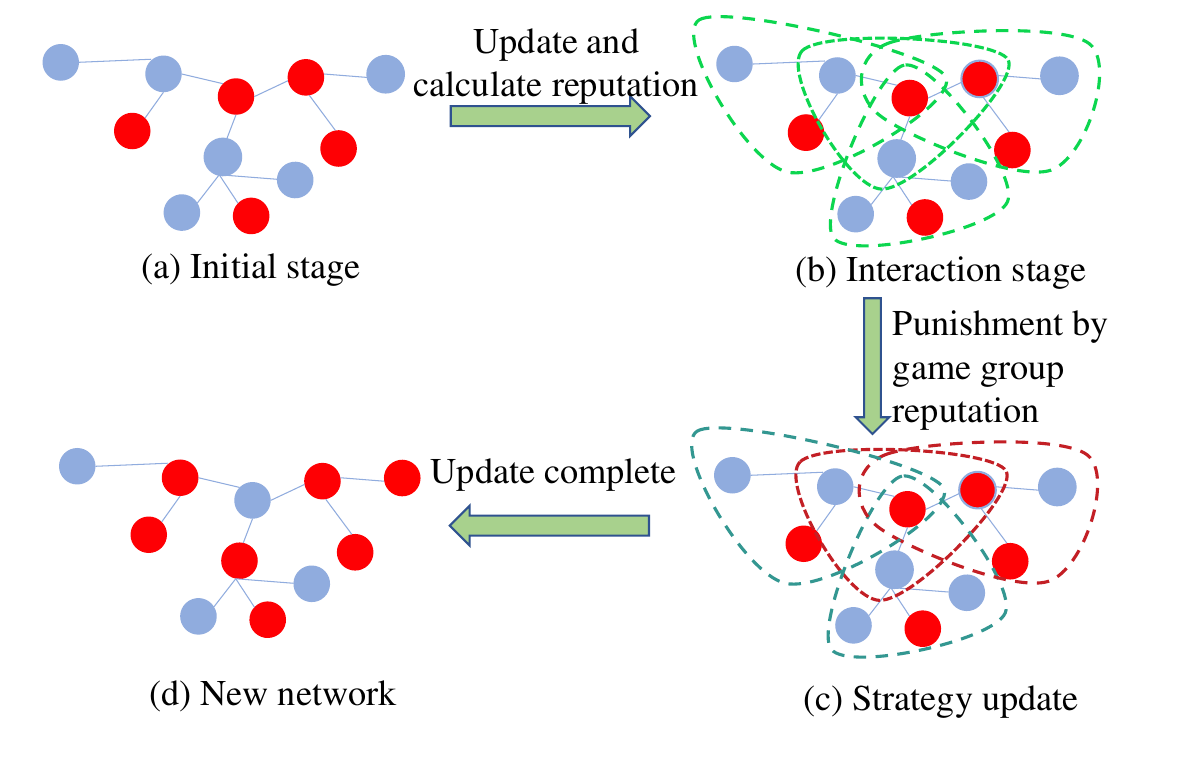}
	\caption{\textbf{The node strategy update process in the network.} Red nodes represent defectors and blue nodes represent cooperators. (a) All nodes are randomly assigned reputations and strategies. (b) All groups of public goods as circled by the dotted line interact. (c) Punishment is carried out according to the average reputation of the game group and update strategy using \cref{eq:4}. Punishment is implemented within the low-reputation group (red dashed box) but not within the high-reputation group (green dashed box). (d) The network is updated, and the next Monte Carlo step is taken.  }
	\label{FIG:1}
\end{figure}

To summarize our model process, \cref{FIG:1} illustrates the process in which nodes in the network start from the initial stage. Their reputation is dynamically updated based on individual strategy, and then the average reputation of the game group is calculated to depend on whether to punish the defector. Finally, the strategy is updated in combination with the new fitness function.

\section{Simulation result and discussion}
\label{simulation}
 This section presents the evolution of cooperation that uses a tolerant punishment mechanism across various parameter settings.  
 
\subsection{Experimental setup and model parameters}

We conducted Monte Carlo simulations on square lattices with periodic boundary conditions, setting $N = 60 \times 60$. The simulation engine was implemented in Python 3.8 within the Anaconda environment. Asynchronous update rules were implemented. Initially, strategies $C$ and $D$ were assigned to each node with the same probability, and each player's reputation was uniformly distributed within the range of $\left [ 0, 20 \right ] $. All simulation evolutionary steps are set to \(t = 10^{4}\). For the final results of the evolutionary process, we ensure the accuracy of the data by averaging 20 independent experiments. Our focus is on the average level of cooperation in the population, defined as $\rho_C = \frac{1}{N}\sum_{i=1}^N s_i$.

\subsection{Integrating reputation and payoff in fitness function}

\begin{figure}[h!]
	\centering
		\includegraphics[scale=.60]{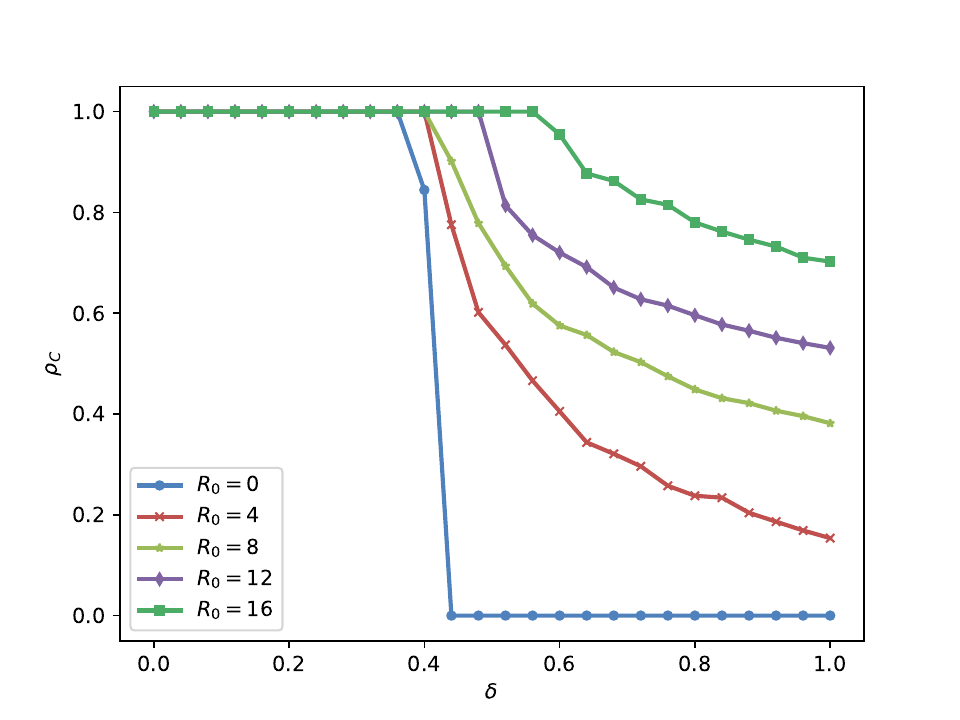}
	\caption{\textbf{Frequency of cooperators $\rho_{C}$ as a function of damping factor $\delta$ for different reputation threshold $R_{0}$}. Each data is obtained by averaging the proportion of cooperators in the last 500 iterations after the system reaches evolutionary stability. Note that the more an individual's fitness depends on reputation, the more it promotes the emergence of cooperation. }
	\label{FIG:2}
\end{figure}

First, we study the change of cooperation ratio $\rho_{C}$  with the damping factor $\delta$ under a different reputation threshold $R_{0}$ to explore how to organically combine reputation and payoff to determine individual fitness, which involves an examination of the evolutionary process of cooperation ratio $\rho_{C}$ as it progresses toward a stable state over time. As illustrated in \cref{FIG:2}, for the cooperation ratio under a different reputation threshold $R_{0}$, when the damping factor $\delta$ is relatively small ($\delta < 0.36$), the cooperation ratio is 1 after the evolution reaches a stable state. That is because when $\rho_{C}$ is relatively small, individual fitness is mainly determined by reputation. Since defection will bring about a reduction in reputation, players will tend to choose cooperation. As time accumulates, cooperators will occupy the entire population.

In addition, as $\delta$ increases, the fitness of the individuals will be primarily influenced by the payoffs. It can be found that for different reputation thresholds $R_{0}$, a phase transition from cooperation to defection will occur, which is due to the fact that the high payoff brought about by free-riding behavior is more conducive to the survival of defectors. In particular, when $R_{0}$ = 0, the cooperation ratio will drop rapidly from 1 to 0. However, when $R_{0} = \{4, 8, 12, 16\}$, due to the emergence of punishment, cooperators can stably appear in the population. It should be noted that when cooperation emerges in the population, the higher the reputation threshold $R_{0}$, the higher the cooperation ratio. This reflects that the looser the inclusion condition of punishment is, the more it can promote cooperation. Based on this simulation, we set $\delta=0.5$ to organically combine reputation and payoff as individual fitness in the following simulations.

\subsection{Influence of enhancement factor and punishment on cooperation}

\begin{figure*}
	\centering
		\includegraphics[scale=.81]{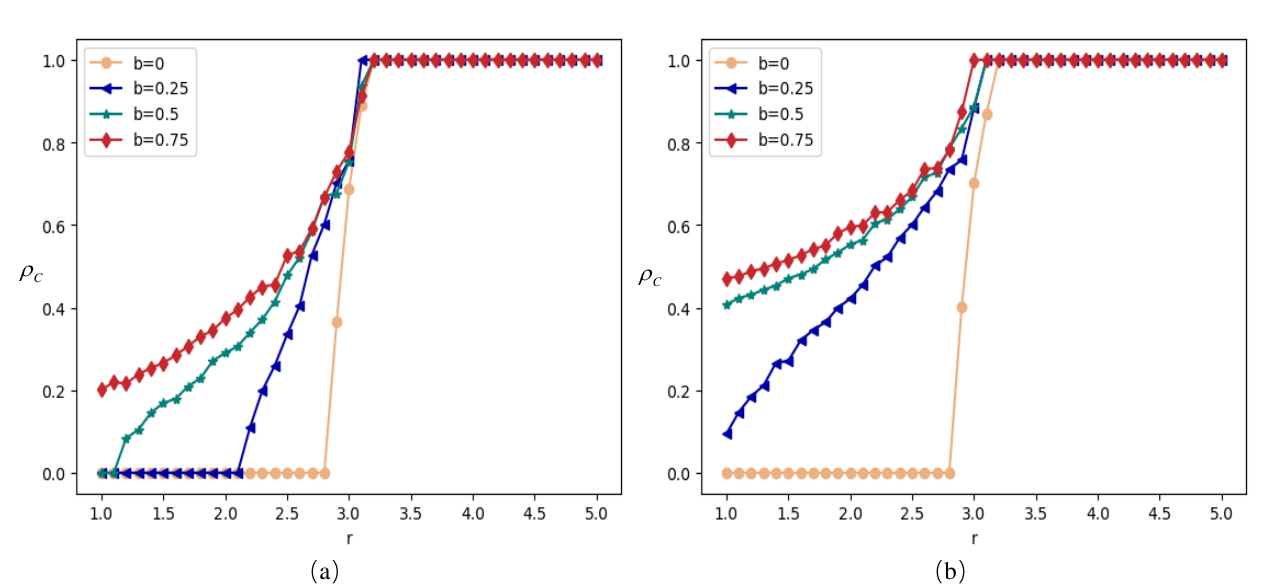}
	\caption{\textbf{Frequency of cooperators $\rho_{C}$ as a function of enhancement factor $r$ for different punishment of $b$}. Each data is obtained by averaging the proportion of cooperators in the last 500 iterations after the system reaches evolutionary stability. Different panels display the cooperation level under different reputation thresholds, as (a) $R_{0}=1$, (b) $R_{0}=4$. The curves show the critical values $r$ at which the phase transition from defection to cooperation occurs under different punishment $b$. For instance, the phase transition points for the emergence of cooperation corresponding to $b$ = 0, 0.25, and 0.50 in (a) are $r$ = 1.1, 2.1, and 2.8, respectively. }
	\label{FIG:3}
\end{figure*}

\begin{figure*}[ht]
	\centering
		\includegraphics[scale=0.45]{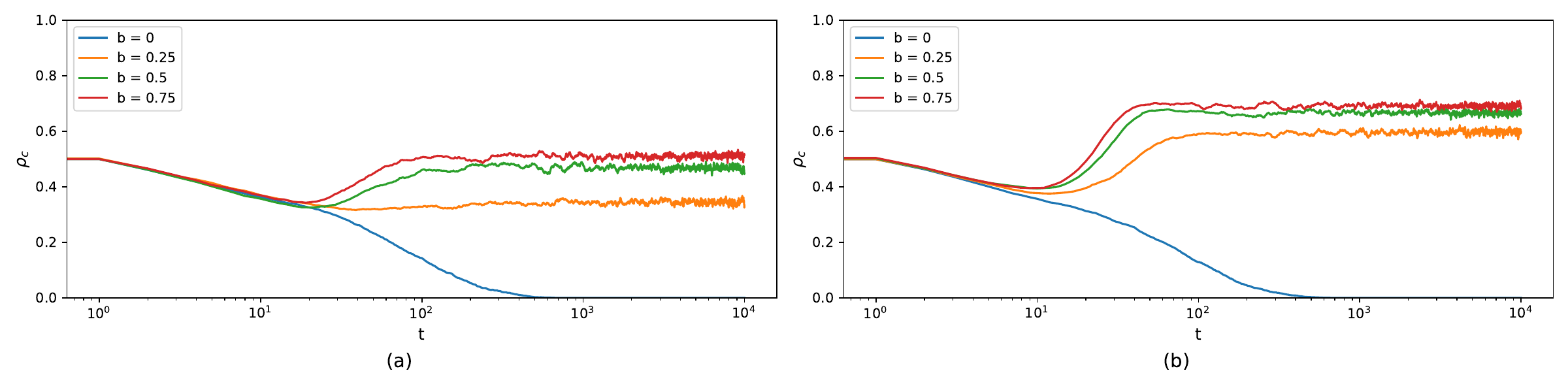}
	\caption{\textbf{Frequency of cooperators $\rho_{C}$ varies with time for different $b$}. Fixed enhancement factor $r$ = 2.5. Different panels display the cooperation level under different reputation thresholds, as (a) $R_{0}=1$, (b) $R_{0}=4$. In different punishment $b$, the system can reach evolutionarily stable in $10^{4}$ iterations. For $b$=0, 0.25, 0.5, and 0.75, the frequency of cooperators initially declines and then gradually increases towards a dynamically stable and non-zero level, whereas $b = 0$, the level of cooperation finally drops to zero. }
	\label{FIG:4}
\end{figure*}

In this section, we analyze the evolution of cooperation under two different reputation thresholds depending on reputation and payoff ($R_{0}=1$ and $4$). From \cref{FIG:3}, when $R_{0}$=1, we observe that for some smaller values of $b$ ($b = 0, 0.25, 0.5$), here $r$ will appear to correspond to a critical value when cooperation arises. When $b=0$, the model degenerates into SPGG with only reputation evolution, and our simulation results are similar to those of the previous study \cite{szolnoki2009topology}. In particular, when $r<2.8$, all players decide to defect, and the ratio of cooperation is 0. When $r=2.8$, a phase transition takes place, leading to an increase in the proportion of cooperators within the population. As the punishment level $b$ increases, the critical value of $r$ required for this phase transition decreases. Especially, when $b=0.75$, the group has a large punishment for the defector. 

Therefore, in this environment with a low reputation threshold and high punishment, defectors are unable to survive, and then cooperation is the dominant strategy. Even when the enhancement factor $r$ is 1, defectors are unable to replace cooperators fully in competition with them either. Hence, a phase transition will not occur. For a more in-depth study, we establish a harsher punishment environment ($R_{0}=4$). Through comparison, we find that in this case only when $b = 0$ will there be a phase transition from defection to cooperation. The result at this time is the same as the result when ($R_{0}=1$). This is because punishment $b=0$. Therefore, regardless of the size of the reputation threshold, there will be no punishment for defectors in the group. When $b$ is equal to 0.25, 0.5, or 0.75, cooperators can survive when $r$ is relatively small, demonstrating that punishment favors cooperation. This shows that punishment under the reputation threshold plays a significant role in the emergence of collective cooperative behavior.

Next, we demonstrate the process of evolution of the cooperation ratio over time under different punishment $b$ with a fixed enhancement factor $r = 2.5$, and in \cref{FIG:4}, we compared two different reputation thresholds ($R_{0}=1$ and $R_{0}=4$).  This value is extremely representative, since cooperation cannot emerge without the introduction of punishment in the SPGG. However, in this model, if reputation and punishment act simultaneously, cooperation can exist under different punishment $b$. As shown in this figure, When in an environment with a relatively large punishment ($b=0.25, 0.50, 0.75$), the cooperation ratio first decreases and then gradually increases to a stable value. After about 100 steps, the proportion of cooperators $\rho_{C}$ in the population does not fluctuate much. This implies that there is intense competition between cooperators and defectors within the population at this time. 

\begin{figure*}
	\centering
		\includegraphics[scale=0.90]{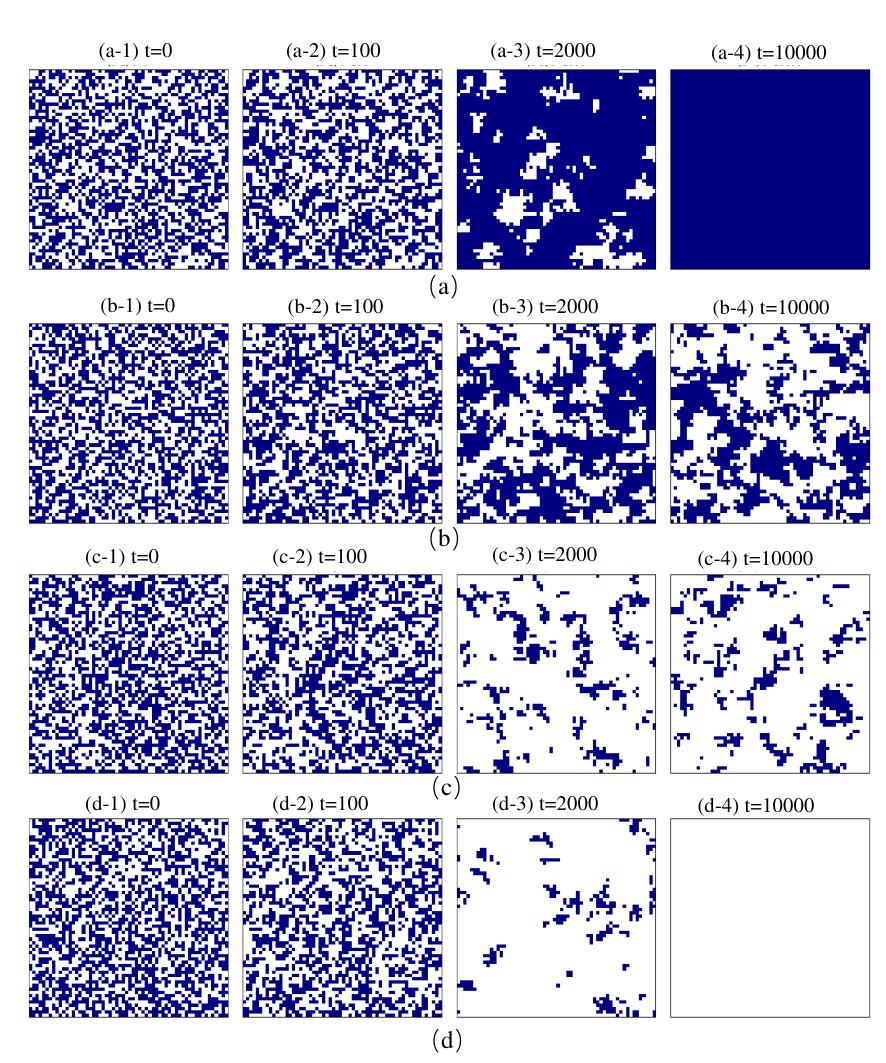}
	\caption{\textbf{Snapshots of the spatial arrangements of strategies at four representative moments for different reputation thresholds.} (a) $R_{0}$= 0; (b) $R_{0}$ = 4; (c) $R_{0}$ = 12 and (d) $R_{0}$=16 and fixed enhancement factor $r$ = 2.5 and $b$=0.2. Each row from top to bottom represents a situation corresponding to a different reputation threshold $R_{0}$. Each column from left to right represents a different time step $t$. White pixels stand for cooperators and blue pixels for defectors. It is noteworthy that defectors can survive in an environment where no punishment is introduced ($R_{0}$ = 0). However, in an environment with a higher reputation threshold ($R_{0}$ = 4, 12, and 16), after 100 time steps, cooperators gradually gather to form stable clusters. Thus, a strict reputation evaluation environment can induce the formation and development of cooperator clusters.}
	\label{FIG:5}
\end{figure*}

\begin{figure*}
	\centering
		\includegraphics[scale=0.95]{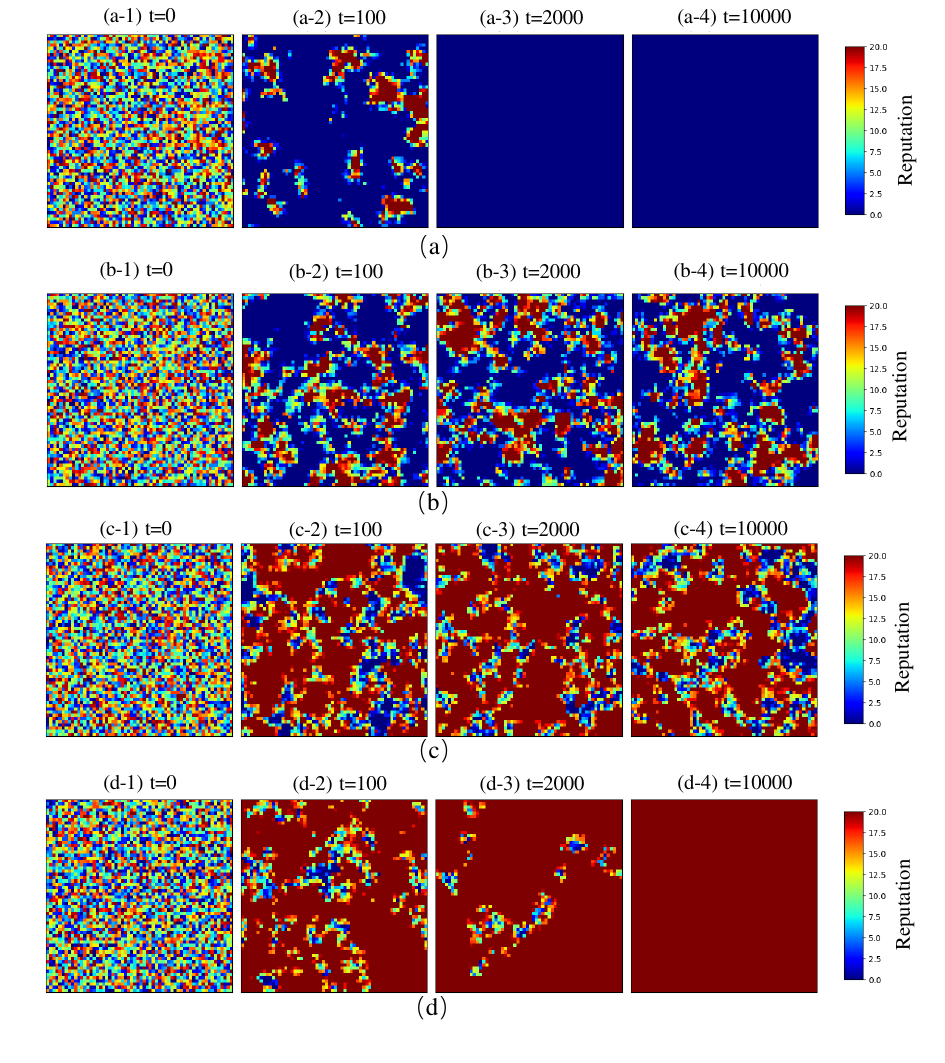}
	\caption{\textbf{Snapshots of the spatial distributions of individuals' reputation at four representative moments for different reputation thresholds}. (a) $R_{0}= 0$; (b) $R_{0} = 4$; (c) $R_{0}=12$ and (d) $R_{0}=16$. Fixed enhancement factor $r=2.5$ and $b=0.2$. From top to bottom, each row corresponds to a situation with a different reputation threshold $R_{0}$. From left to right, each column represents a different time step $t$. The color bar shows the reputation values corresponding to different colors. It should be noted that the reputation distribution snapshots are not entirely in line with the strategy distribution snapshots in \cref{FIG:5} because the reputation evaluation criteria vary under different reputation thresholds.}
	\label{FIG:6}
\end{figure*}

However, in the least favorable environment ($b=0$), that is, regardless of the reputation threshold, there will be no
punishment for defectors in the environment. Eventually, the proportion of cooperation will drop to zero. That is because when $b = 0$, the game degenerates into a PGG with only a reputation mechanism. At this time, the strategy update rule of the nodes depends on the income difference and their reputation. The income difference becomes the factor that dominates strategy updates so that cooperators will tend to imitate the strategies of defectors, resulting in free-riding behavior and leading to the cooperation proportion dropping to zero. When $b$ = 0.25, 0.50 and 0.75, punishment plays a role in promoting cooperation in the cluster. Therefore, the cooperation level reaches a dynamic equilibrium and is in a mixed state of cooperators and defectors. By comparing these four curves, it can be found that there is a risk of inducing cooperators to become defectors. Punishment below the reputation threshold can maintain the proportion of cooperation at a medium level. In addition, in the groups with the highest punishment $b$, cooperators have more advantages. This consolidates the conclusion in \cref{FIG:1} that the greater the punishment, the more it can promote cooperation in SPGG. Moreover, by comparing the cooperation evolution with different reputation thresholds ($R_{0}=1$ and $R_{0}=4$), it can be found that a higher reputation threshold ($R_{0}=4$) leads to a stricter environment of punishment, that is, in the case of the same punishment $b$, a higher reputation threshold results in a wider range of punishment, thereby enhancing the survival of cooperators, which is consistent with the result shown in \cref{FIG:3}.

\subsection{Evolution of strategy and reputation on the lattice}

\begin{figure*}[ht]
	\centering
		\includegraphics[scale=1.0]{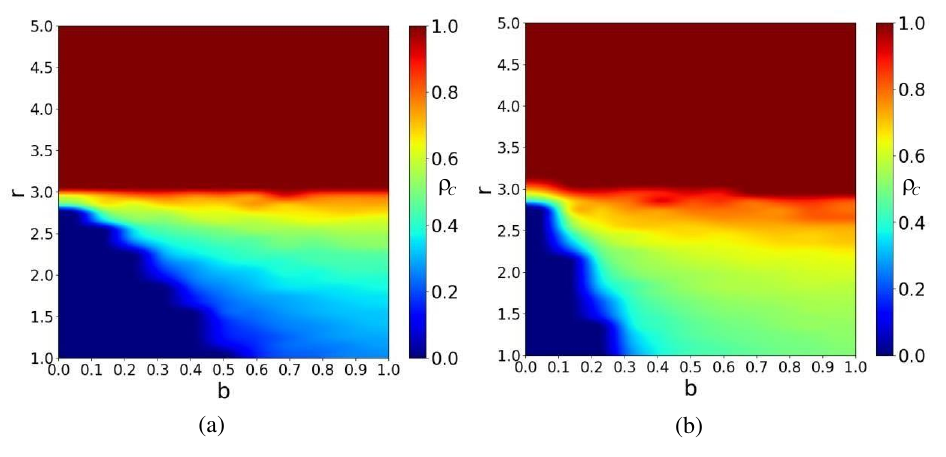}
	\caption{\textbf{Heatmap of cooperation under different reputation thresholds}.  Different panels display the cooperation level under different reputation thresholds, as (a) $R_{0}=1$, (b) $R_{0}=4$. The density $\rho_{C}$ of cooperators is shown in the $r-b$ parameter plane. The legend on the right side of the panel explains the meaning of colors. The value of the cooperation level $\rho_{C}$ is the average of the last 500 steps obtained through 10,000 MC steps. It is generally the case that when we increase the factor $r$ or penalty $b$ simultaneously, the level of cooperation will also increase correspondingly. However, when $r$ reaches a certain threshold, regardless of how large the punishment $b$ is, pure cooperation will appear in the entire population.}
	\label{FIG:7}
\end{figure*}
To gain deeper insights into the emergence of cooperation and the formation of cooperative clusters in space, we analyze the spatial distribution of defectors and cooperators as shown in \cref{FIG:5}. This figure illustrates the evolutionary dynamics of the spatio-temporal distribution of both strategies at different time intervals. At the beginning ($t\le100$), the node strategies are randomly distributed. We observe that only a small number of nodes are clustered. This is possibly due to the random distribution of node strategies and reputations, which means that cooperators are unable to identify defectors well in the group. Therefore, cooperators find it difficult to resist the invasion of defectors, causing defectors to occupy the most population. As time passes, the role of reputation and punishment becomes more and more obvious, providing a good environment for the survival of cooperators. 

In the dual mechanism of reputation and punishment ($R_{0}=4,12,16$), as the simulation evolves, the role of reputation and punishment becomes more and more obvious, providing a good environment for the survival of cooperators. With the survival of the cooperators, the remaining cooperators form clusters to resist the invasion of defectors. At time step $t = 10000$, cooperators will dominate, especially when $R_{0}=16$, cooperators will completely replace defectors. However, when $R_{0}=0$, at this time, the average reputation of all groups will be higher than this reputation threshold, resulting in the inability to incorporate punishment. Therefore, cooperators cannot resist the invasion of defectors, and thus there is the behavior of all players free-riding in the end. The results indicate that a lenient environment is not suitable for facilitating cooperation or resisting the invasion of defectors. Eventually, cooperators in the population completely disappear. In contrast, a strict environment, that is, the higher the reputation threshold, the more it can promote the emergence of cooperation.

Similarly, based on the snapshots of strategies shown in \cref{FIG:5}, we have represented snapshots of individual reputation distributions in \cref{FIG:6} to more thoroughly study the influence of reputation on the evolutionary game. In various scenarios, initial node reputation distributions are uniformly distributed within the range of [0, 20]. As demonstrated in \cref{FIG:6}, in groups where reputation and punishment act in concert ($R_{0}=4,12$), at $t=100$ individuals with high reputation will gradually emerge. Thus, a player with high reputation will exert an impact on the strategy update of those with low reputation, making the latter more inclined to learn the cooperative strategy of the former, thus leading to an increase in their own reputation. Consequently, at $t=10000$, individuals with high reputation will gradually form groups, thus resisting the invasion of the defection strategy of individuals with low reputation, and the phenomenon of high reputation being surrounded by low reputation will occur. 

It should be noted that when $R_{0}=0$, the snapshot of the reputation evolves to pure blue, which means that the reputation corresponding to its players is the lowest reputation. This is because players in the group will adopt the more profitable defection behavior because they will not be punished for defection, resulting in all players in the group choosing to defect, thus low-reputation individuals occupying the entire population. However, when $R_{0}=16$, At $t = 100$, a large number of players with a high reputation will occupy the entire population. At $t = 10000$, they will completely occupy it. This is because when the reputation threshold is set too high, it will severely crack down on most defecting individuals. As a result, the group is more inclined to choose the cooperative strategy. Moreover, due to cooperation, the reputation of players accumulates rapidly. Thus, high-reputation players become part of the entire population.

\subsection{Impact of reputation-punishment synergy on cooperation}
Finally, as illustrated in \cref{FIG:7}, we study how the level of cooperation changes by varying the main parameters of the model ( the enhancement factor $r$ and the punishment $b$) under the extended model of cluster reputation punishment. To gain a generally valid observation on the behavior of the system, we present the results obtained simultaneously under different reputation thresholds ($R_{0}=$1 and 4). Under both reputation thresholds, the density of cooperation increases as the parameter $ r $ or $ b $ increases. Within the explored range of parameters, regions of pure cooperation and pure defection are observed. In general, higher values of $ r $ and $ b $ lead to a higher likelihood of cooperation, while lower values are associated with an increase in defection. Moreover, when the enhancement factor $ r $ reaches a certain threshold, a pure cooperation region emerges, regardless of the magnitude of the third party punishment $ b $. In addition, when $r$ is sufficiently large, the proportion of cooperators in the system changes less significantly with increasing $b$. This can be attributed to the increase in $r$, which leads to a significant growth in the payoffs of both cooperators and defectors. Consequently, the impact of the payoff difference caused by punishment on the change in the proportion of cooperation is reduced. In this kind of SPGG with high benefits, players are more inclined to choose the cooperation strategy. By comparing the heatmaps with two different reputation thresholds, it can be observed that when $R_{0}=4$, which corresponds to a higher reputation threshold, the cooperative region is larger than that of a lower reputation threshold ($R_{0}=1$). Similarly, the defective region is also smaller. This further confirms the synergistic effect of reputation and punishment in promoting cooperation in grid networks. This aligns with the theoretical expectations of the extended public goods game presented in the paper.

\section{Conclusions and outlook}\label{outlook}

In this paper, we enhance the existing SPGG from two perspectives. First, we consider the average reputation of the game group and the reputation threshold in reputation evolution to determine punishment. Second, we assume that players with high payoff and reputation have a higher probability of learning strategies. Therefore, the paper combines reputation and punishment to jointly construct the player fitness function and extends the individuals with a higher reputation threshold who are fitter to the traditional strategy update. A reputation threshold was included into the model to distinguish between the high-reputation and low-reputation groups of the game, thus determining whether to punish during the game interaction process. 

We have examined the effects of an enhanced reputation threshold in SPGG, along with a punishment mechanism that incorporates tolerance and strategy learning influenced by both income and reputation on the evolution of cooperation. Our simulations demonstrated that incorporating reputation into the fitness function is crucial to fostering cooperation, particularly in stringent environments. Generally, imitation based on income and reputation, combined with a rigorous reputation evaluation framework, supports the sustainable survival of cooperators. This mechanism provides valuable information on the ubiquitous phenomenon of cooperation within social systems.

There are still many shortcomings in our model that need to be improved. For example, when making the decision on whether to punish or not, the reputation threshold in our model is fixed for all game groups. This fact simplifies real situations, more complex in general. In addition, when constructing the fitness function using payoff and reputation, we use a simple linear function to combine both reputation and payoff. Other extensions can also be used in more in-depth research on the evolution of cooperation in SPGG. For example, certain rewards can be given for cooperative behaviors in low-reputation groups. Punishment and reward values can be linked to individual reputations rather than a fixed value. 

\section*{ACKNOWLEDGEMENTS}

Gui Zhang, Y. Yao, Z. Zeng and M. Feng are supported by grant No. 62206230 funded by the National Natural Science
Foundation of China (NSFC), and grant No. CSTB2023NSCQ-MSX0064 funded by the Natural Science
Foundation of Chongqing. M. Chica is supported by EMERGIA21\_00139, granted by Consejería de Universidad, Investigación e Innovación of the Andalusian Government.

\bibliographystyle{unsrtnat}
\bibliography{aipsamp}

\end{document}